\pdfoutput=1
\documentclass[twocolumn, prb, showpacs,preprintnumbers,amsmath,amssymb]{revtex4}

\usepackage{graphicx}
\usepackage{bm}
\usepackage{epsfig}
\usepackage{dcolumn}
\newcommand{\nbf}{{NbFe$_{2}$}}
\newcommand{\nbfy}{{Nb$_{1-y}$Fe$_{2+y}$}}

\begin{document}

\preprint{}

\title{Doping driven magnetic instabilities and quantum criticality of NbFe$_{2}$}

\author{D. A. Tompsett}
\email{dat36@cam.ac.uk}
\author{R. J. Needs}
\author{F. M. Grosche}
\author{G. G. Lonzarich}


\affiliation{%
Cavendish Laboratory, University of Cambridge, JJ Thomson Ave., Cambridge CB3 0HE, UK}%

\date{\today}

\begin{abstract}
Using density functional theory we investigate the evolution of the magnetic ground state of NbFe$_{2}$ due to doping by Nb-excess and Fe-excess. We find that non-rigid-band effects, due to the contribution of Fe-\textit{d} states to the density of states at the Fermi level are crucial to the evolution of the magnetic phase diagram. Furthermore, the influence of disorder is important to the development of ferromagnetism upon Nb doping. These findings give a framework in which to understand the evolution of the magnetic ground state in the temperature-doping
phase diagram. We investigate the magnetic instabilities in NbFe$_{2}$. We find that explicit calculation of the Lindhard function, $\chi_{0}(\mathbf{q})$, indicates that the primary instability
is to finite $\mathbf{q}$ antiferromagnetism driven by Fermi surface nesting. Total energy calculations indicate that $\mathbf{q}=0$ antiferromagnetism is the ground state. We discuss the influence of competing $\mathbf{q}=0$ and finite $\mathbf{q}$ instabilities on the presence of the non-Fermi liquid behavior in this material.
\end{abstract}

\maketitle

\section{\label{sec:H1}Introduction\protect}
Materials that exhibit magnetic phase transitions that may be tuned to zero temperature are strong candidates for the formation of novel ordered states around the quantum critical point. These novel states rely upon the breakdown of assumptions that underpin the Fermi liquid theory of metals when the magnetic interactions become extended in space and time. Such a situation is possible at a magnetic quantum critical point and may lead to the formation of new phases of matter that are potentially technologically valuable. The formation of superconductivity in cuprates and iron-pnictides near magnetic phase transitions are among the phenomena that motivate us to understand materials where magnetic interactions are potentially of crucial importance.

The metallic compound \nbf~has been of interest to researchers for decades due to its rich magnetic phase diagram. Initially, stoichiometric \nbf~was thought to be either paramagnetic or ferromagnetic\cite{NevittNb1, YamadaNb3}. However, subsequent NMR\cite{YamadaNb2} and magnetization measurements\cite{YamadaNb1} have been interpreted as showing spin density wave (SDW) order with a N\'{e}el temperature $T_{N}\approx10K$. 

Studies of the doping evolution of the phase diagram have shown that this spin density wave state is very sensitive to changes in stoichiometry. Studies of doping in \nbfy~with either Nb-excess ($y < 0$) or Fe-excess ($y > 0$) have shown that stoichiometric \nbf~becomes ferromagnetic upon either electron or hole doping\cite{Crook1}. Because of the close proximity of the reported SDW and ferromagnetism, \nbf~has long been considered a candidate material for the coexistence of SDW and ferromagnetic fluctuations\cite{Shiga1,Crook1}.

\begin{figure}
\centering
\begin{tabular}{c}
\includegraphics[width=0.97\linewidth,angle=0]{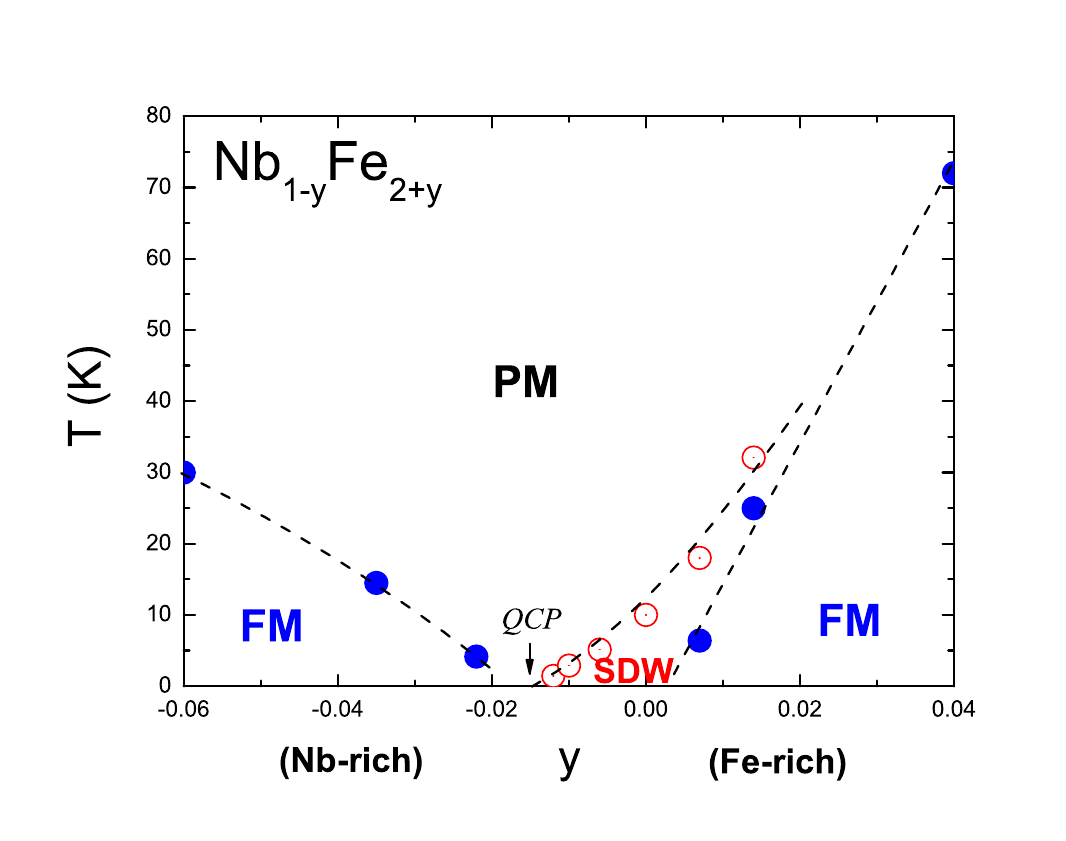} \\
\end{tabular}
\caption{\label{fig:DNbFe2} (Color online) Phase diagram for \nbfy\cite{Moroni1}. By adjusting the precise composition
within a narrow range, \nbf~can be tuned from
ferromagnetism (FM) ($y > 0.01$) via an intermediate spin-density-wave (SDW) modulated state to a quantum critical point
(QCP) ($y \approx -0.015$).}
\end{figure}

Recent experimental work has further probed the behavior of the temperature-doping phase diagram of high purity \nbf~via measurements of the temperature dependence of resistivity, specific heat and magnetic susceptibility\cite{Brando1, Moroni1}. The resulting phase diagram is shown in Fig.~\ref{fig:DNbFe2}. Importantly these refined measurements indicate the presence of a quantum critical point as the reported SDW collapses when \nbf~is tuned away from stoichiometry by Nb-excess. At this critical point, intriguing non-Fermi liquid behavior in the material has been demonstrated. Wada \textit{et al.}\cite{WadaNb1} had previously reported an enhanced electronic specific heat in the late 1980's. This behavior has been confirmed by recent measurements\cite{BrandoNb2,BrandoNb3,Brando1} that describe the non-Fermi liquid behavior in more detail. They show at low temperatures near the quantum critical point the presence of a logarithmic divergence
of the specific heat and a $\rho \propto T^{\frac{3}{2}}$ dependence in the resistivity.

Further NMR and magnetization studies have confirmed the presence of antiferromagnetism at stoichiometry\cite{YamadaNb4, Crook1, KurisuNb1}, but neutron diffraction\cite{BrandoNb3, Moroni1} studies have been unable to determine the magnetic order. Therefore, studying the electronic and magnetic properties of this material with electronic structure calculations may aid our understanding and has been addressed previously\cite{Ishida1, Asano1, Asano2}. Total energy calculations with the frozen core approximation\cite{Asano1} suggested a paramagnetic ground state, while work with the LMTO method with shape approximations\cite{Asano2} suggested an antiferromagnetic ground state nearly degenerate with a paramagnetic state. 

Recent work by Subedi \textit{et al.} has discussed in detail the magnetic interactions of local moment configurations within the unit cell\cite{SubediNb1} of stoichiometric \nbf. \nbf~exhibits significant exchange splitting of the 3$s$ core levels\cite{vanAckerNb1}. This indicates the presence of local moment fluctuations as is also found in the Fe-based superconductors\cite{BondinoNb1}. Subedi \textit{et al.} found that the magnetism has itinerant character due to the dependence of the size of the moments on the magnetic configuration. Also, a ferrimagnetic ground state was found to be most favourable. Furthermore large moments $\sim 1 \mu_B$ were obtained, which far exceed the $\sim0.02 \mu_B$ inferred from experiment. In this study we further address the magnetic ground state of this material using the full potential methods in WIEN2k\cite{Blaha1}. We consider the doping dependence of the ground state and discuss the magnetic interactions that may be key to the critical behavior in this material. Our findings for the ground state are used to discuss the driving forces behind the rich magnetic phase diagram and the role of magnetism in the intriguing non-Fermi liquid behavior of this material.

We begin our analysis by considering the electronic structure of stoichiometric \nbf. All calculations are performed using the GGA-PBE\cite{PerdewNb1} correlation functional of Perdew, Burke and Ernzerhof. We have used the experimental lattice parameters\cite{Moroni1}. For stoichiometric \nbf~$a$ = 4.8401(2) \AA~and $c$=1.8963(6) \AA. All results presented were obtained using $RK_{max}=8$, $G_{max}=16$ and IFFT factor=4.0. Here, $RK_{max}$ determines the matrix size ($K_{max}$ is the
plane wave cut-off, $R$ is the smallest of the atomic sphere radii), $G_{max}$ is the maximum wavevector in the charge density Fourier expansion and IFFT (Indices of the Fast Fourier Transfrom grid) determines the size of the mesh for the calculation of the exchange-correlation potential in the interstitial region. The Fermi surface and charge densities were calculated from a 43$\times$43$\times$23 grid of k-points. The radii of the muffin tins were 2.12$a_0$ for Fe and 2.15$a_0$ for Nb.

\section{\label{sec:H1}Crystal and Bonding Structure\protect}

\begin{figure}
\centering
\includegraphics[width=0.25\textwidth,angle=0]{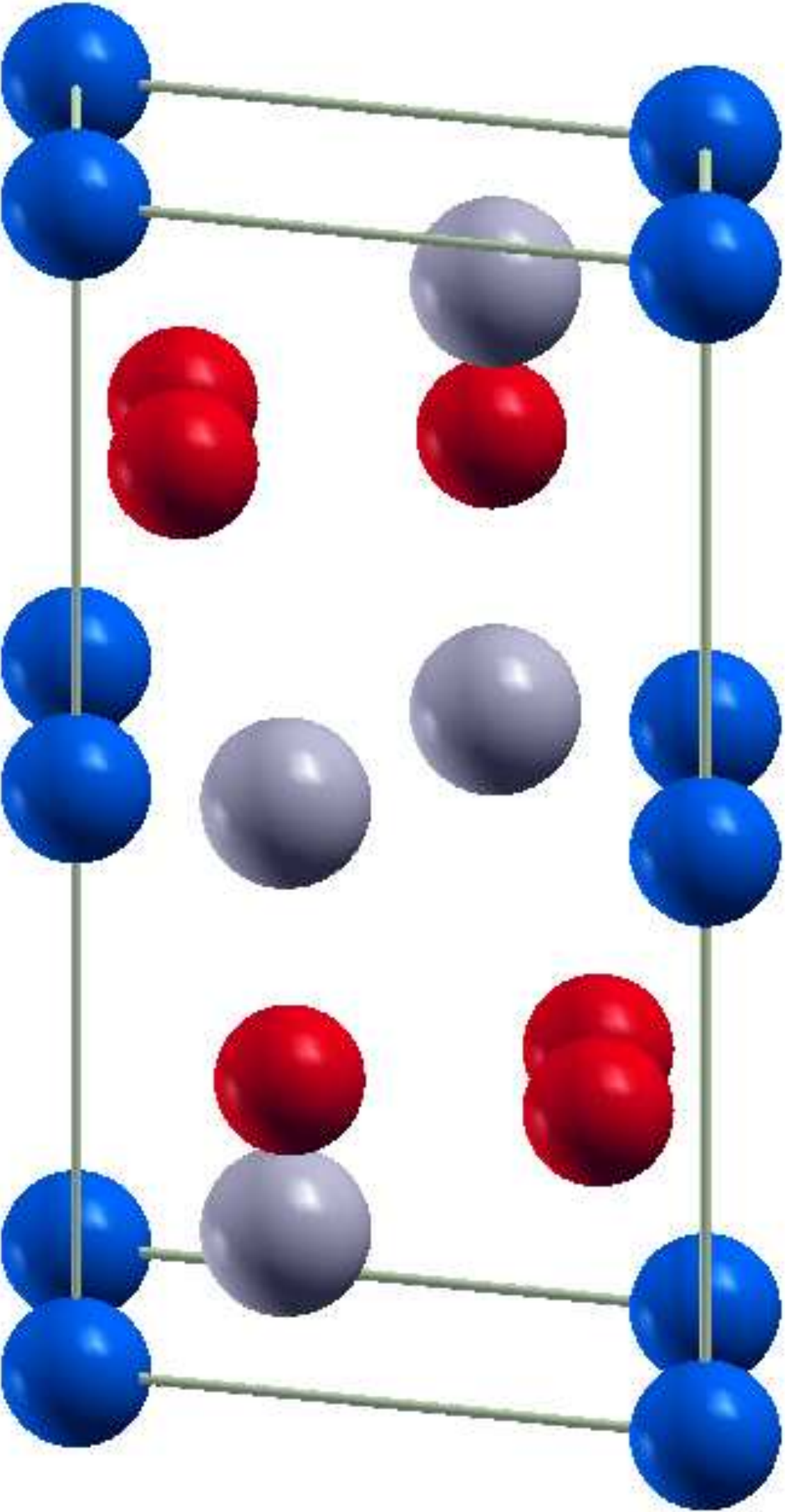}

\caption{\label{fig:NbFe2Struct} (Color online) Crystal Structure of \nbf. Fe(6\textit{h}) red, Fe(2\textit{a}) blue and Nb grey.}
\end{figure}

\nbf~crystallizes in the C14 Laves phase which has the MgZn$_{2}$ hexagonal structure. The crystal structure is shown in Fig.~\ref{fig:NbFe2Struct}. The Fe atoms form a layered structure of a kagome lattice (6\textit{h} sites) separated by Fe (2\textit{a} sites) atoms centred on the line between alternate kagome triangles. The Nb atoms occupy the interstices in this Fe structure and lie slightly out of plane with respect to the Fe 2\textit{a} sites. The site symmetries have internal degrees of freedom: Nb at 4$f$(1/3/,2/3,x) and Fe at 2$a$(0,0,0) and 6$h$(y,2y,3/4). In our calculation of stoichiometric \nbf~we have relaxed these internal parameters resulting in x=0.0657 and y= 0.1705. 

\begin{figure}
\centering
\begin{tabular}{cc}
\includegraphics[width=0.27\textwidth,angle=90]{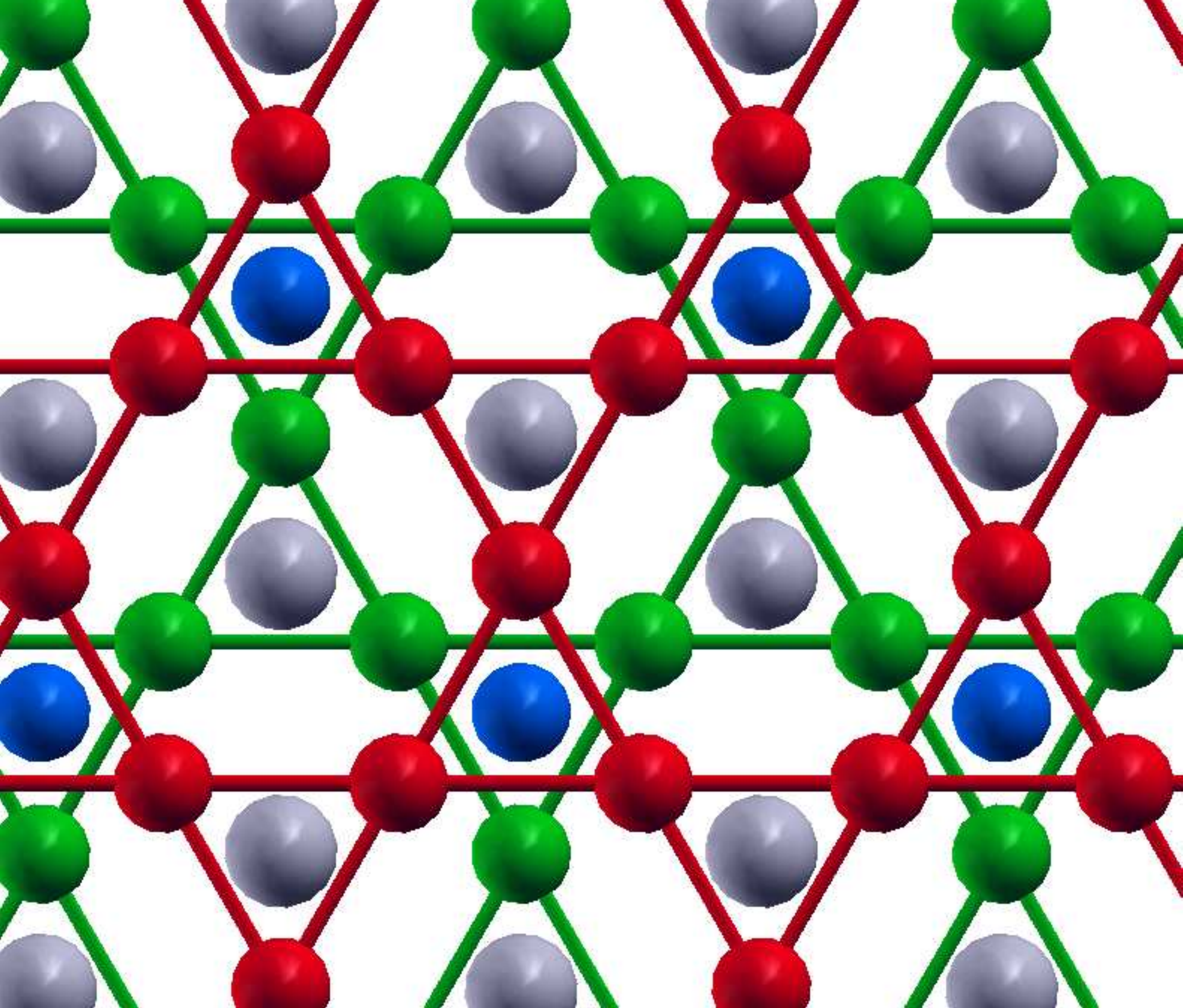} &
\includegraphics[width=0.27\textwidth,angle=90]{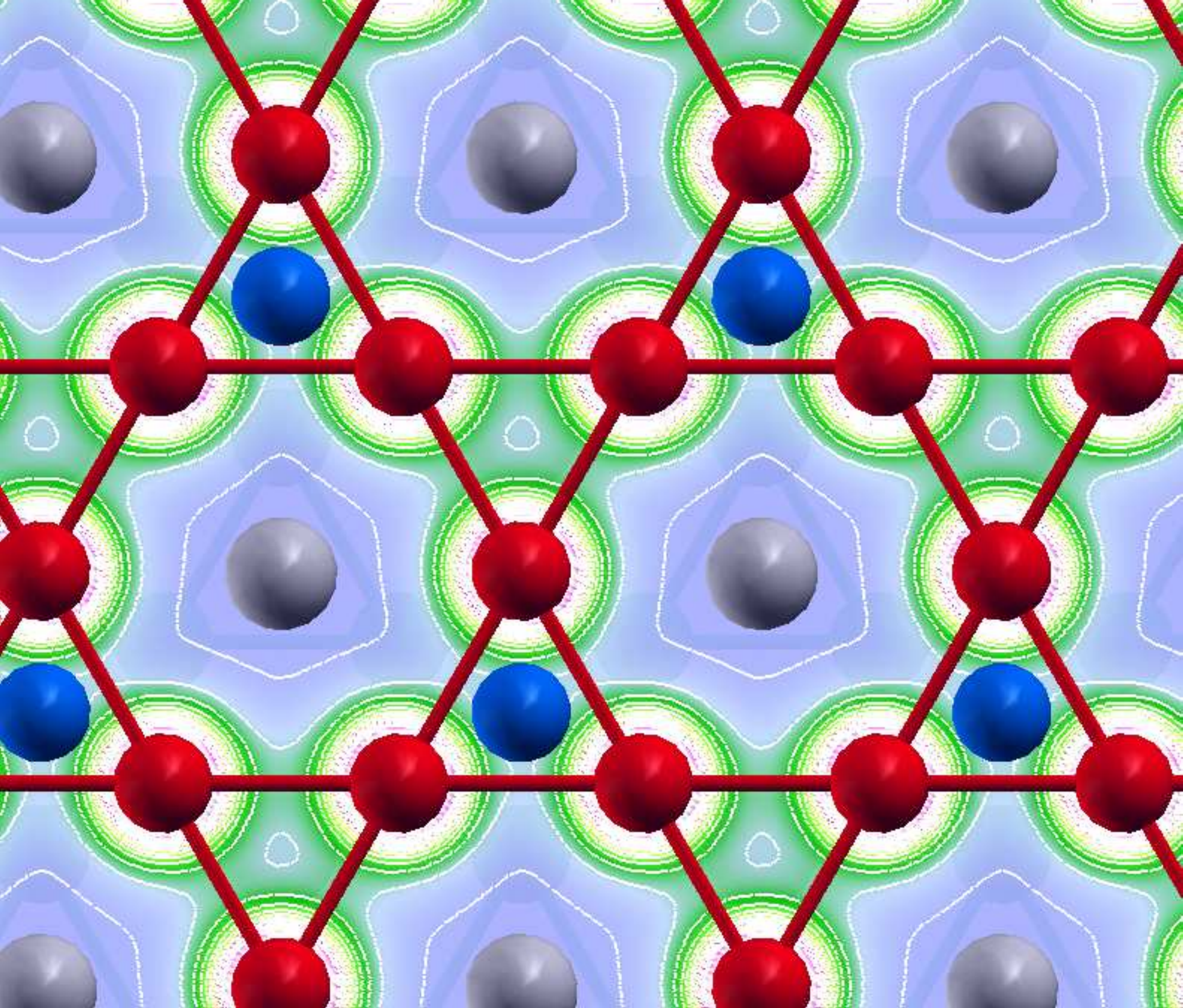} \\
\end{tabular}
\caption{\label{fig:NbFe2Struct2} (Color online) Crystal Structure of \nbf~at left. View shown along the \textbf{c}-axis. The upper and lower kagome layers formed by the Fe(6\textit{h}) sites have been separated into red and green respectively. Fe(6\textit{h}) red (upper kagome), Fe(6\textit{h}) green (lower kagome), Fe(2\textit{a}) blue and Nb grey. In the right figure we show charge density contours in the upper kagome plane. The charge density for kagome triangles coordinated with Fe(2\textit{a}) sites is inequivalent to those with no coordination.}
\end{figure}

It is worth considering this crystal structure in further detail since its properties affect the bonding, doping impurity location and the magnetic coupling. In Fig.~\ref{fig:NbFe2Struct2}(a) we show an extended view of the crystal structure along the \textbf{c}-axis. We can see that the blue Fe(2\textit{a}) atoms only coordinate every second triangle of the kagome structure formed by the Fe(6\textit{h}) sites. To emphasise this we have separated the colors of the upper and lower kagome layers into red and green. The presence of the Fe(2\textit{a}) atoms coordinating every second kagome triangle means that the bonding between the pairs of Fe(6\textit{h}) atoms in the kagome layers will not be the same. In Fig.~\ref{fig:NbFe2Struct2}(b) we show charge density contours within the kagome layer. As we can see the bonding between Fe(6\textit{h}) sites varies significantly. Effectively the presence of the Fe(2\textit{a}) sites draws away charge density into the out of plane Fe(2\textit{a})-Fe(6\textit{h}) bonds, and starves the bonds in the kagome triangles that are coordinated with an Fe(2\textit{a}) site.



In Table~\ref{tab:NbFe2BondLengths} we show the interatomic bond distances. Significantly, amongst the Fe sites, the nearest neighbour distances from Fe(2\textit{a}) and Fe(6\textit{h}) sites are similar. Therefore upon Nb doping, the preferred dopant site is likely to be
determined by the bonding network of the Fe cage rather than the
volume available at the site. We investigate this via total energy calculations in the following section.

\begin{table}
\caption{\label{tab:NbFe2BondLengths} Interatomic bond distances for stoichiometric \nbf.
In each case we show the shortest distance between the two given sites.}
\begin{ruledtabular}
\begin{tabular}{cc}
 Sites & Distance (\AA) \\
\hline
Fe(2\textit{a})-Fe(6\textit{h}) & 2.42\\
Fe(6\textit{h})-Fe(6\textit{h}) & 2.37\\
Fe(2\textit{a})-Fe(2\textit{a}) & 3.95\\
Nb-Fe(2\textit{a}) & 2.84\\
Nb-Fe(6\textit{h}) & 2.81\\
Nb-Nb & 2.89\\
\end{tabular}
\end{ruledtabular}
\end{table}

\begin{figure}
\centering
\begin{tabular}{c}
\includegraphics[width=0.5\textwidth,angle=0]{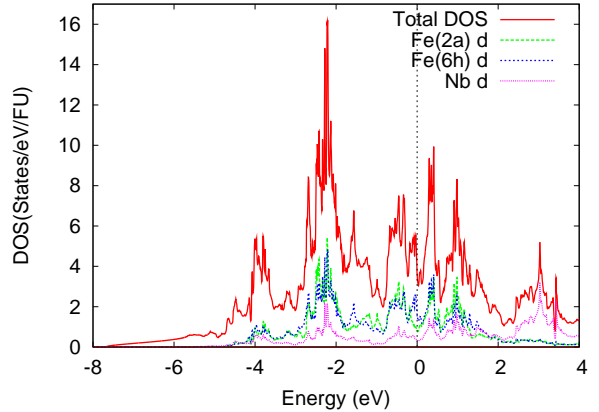} \\
\end{tabular}
\caption{\label{fig:DOSNbFe2NM} (Color online) Non-magnetic calculated DOS for NbFe$_{2}$. The Fermi level is set to 0 eV.}
\end{figure}

The density of states for non-magnetic \nbf~is shown in Fig.~\ref{fig:DOSNbFe2NM} and agrees with the results of previous studies\cite{Ishida1,SubediNb1}. We also show the partial density of states by atomic site. The partial density of states for Fe(2\textit{a})-\textit{d} and Fe(6\textit{h})-\textit{d} sites possess similar structure, which is consistent with their strong bonding exhibited in the charge density. The Fe-\textit{d} states dominate the structure at the Fermi level. In contrast the Nb-\textit{d} character  contributes far less structure to the overall density of states. This reflects the role of Nb as an electron donor in the system.

Clearly the Fermi level lies near a depression in the density of states which has been cited as important to the formation of a SDW at stoichiometry\cite{SubediNb1}. The density of states at $E_F$ is 3.6 eV$^{-1}$FU$^{-1}$. In a rigid-band approximation the impact of doping a small amount of holes (e.g. Nb-doping) will raise the total density of states and promote the development of ferromagnetism as is seen in its doping phase diagram experimentally\cite{Brando1}. However, if we take the same rigid-band approach to electron doping, such as in the Fe-doped system, then the density of states is expected to fall. This is in conflict with the experimental phase diagram (see Fig.~\ref{fig:DNbFe2}), which shows the formation of ferromagnetism with small amounts of Fe doping. Therefore, effects beyond rigid-band shifts of $E_F$ may be in operation and we attempt to explore these in the following section.

\section{\label{sec:Doping}Doping Dependence of the Ferromagnetic Instability\protect}
We may evaluate the proximity of the system to ferromagnetism within Stoner theory\cite{StonerNb1}, where if $N(E_F)I > 1$ then a material is unstable to ferromagnetism. It has been shown\cite{SubediNb1} that if we assume a typical interaction\cite{AndersenNb1,PoulsenNb1} $I \sim 0.7 - 0.9$ eV then we obtain a Stoner enhancement of $(1-N(E_F)I)^{-1} \sim 3$ for stoichiometric \nbf. Experimentally\cite{BrandoNb3,Brando1}, however, comparing the measured susceptibility with the bandstructure value of $N(E_F)$ indicates a Stoner enhancement $(1-N(E_F)I)^{-1} > 100$. This suggests firstly, that the interaction scale may be unusually large in \nbf, $I > 1$ eV. Secondly, very slight increases in the value of $N(E_F)$ of order a few percent should drive \nbf~to a ferromagnetic instability. In this section we utilise doped supercell calculations to investigate the evolution of $N(E_F)$ with both hole (Nb) and electron (Fe) doping. We note that doped \nbf~samples are disordered as is evidenced by the significant increases in the low temperature resistivities that occur upon doping\cite{Moroni1}. For these calculations we have employed the lattice parameters derived from experiment\cite{Moroni1} and relaxed all internal coordinates. We consider the non-magnetic density of states in order to apply the Stoner framework.

%
%
%

First, we consider hole doping on the Nb-rich side of the phase diagram of \nbfy. We consider doping corresponding to $y=-0.065$ Nb-rich Nb$_{1.065}$Fe$_{1.935}$. This state is formed by substituting one Nb atom for an Fe atom in  a $2 \times2$ supercell and lies beyond the formation of ferromagnetism in the experimental phase diagram. If we were to apply the rigid band approximation to the density of states shown in Fig.~\ref{fig:DOSNbFe2NM}, then we expect that for Nb$_{1.065}$Fe$_{1.935}$ the density of states at the Fermi level, $N(E_F)$, will rise to 5.4 eV$^{-1}$FU$^{-1}$. Therefore the system would satisfy the Stoner criterion, $N(E_F)I > 1$, for ferromagnetism. Clearly, since there are the two inequivalent Fe(2\textit{a}) and Fe(6\textit{h}) sites, then the Nb might be doped at either one. We have performed supercell calculations with the Nb atom substituted at both sites. Non-magnetic total energy calculation for the $2 \times2$ supercell indicated that for substitution at the Fe(6\textit{h}) site the total energy is $\sim 6.8$eV/FU lower than that when we substitute at the Fe(2\textit{a}) site. Therefore, the total energy strongly favours substitution at the Fe(6\textit{h}) site.

\begin{figure}
\centering
\begin{tabular}{c}

(a) Nb$_{1.065}$Fe$_{1.935}$ \\
\includegraphics[width=0.5\textwidth,angle=0]{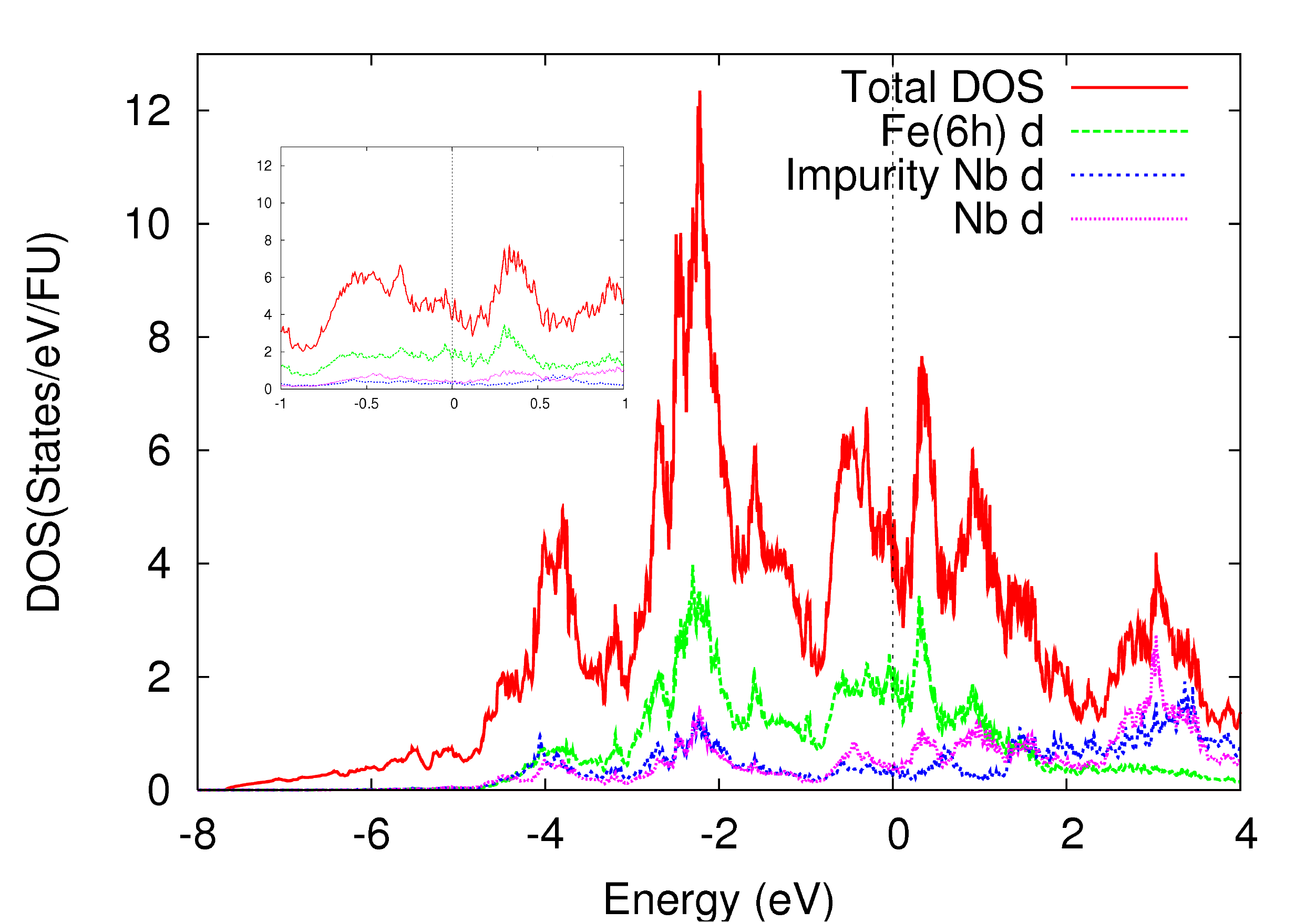}
\\
\\
(b) Nb$_{0.935}$Fe$_{2.065}$ \\
\includegraphics[width=0.5\textwidth,angle=0]{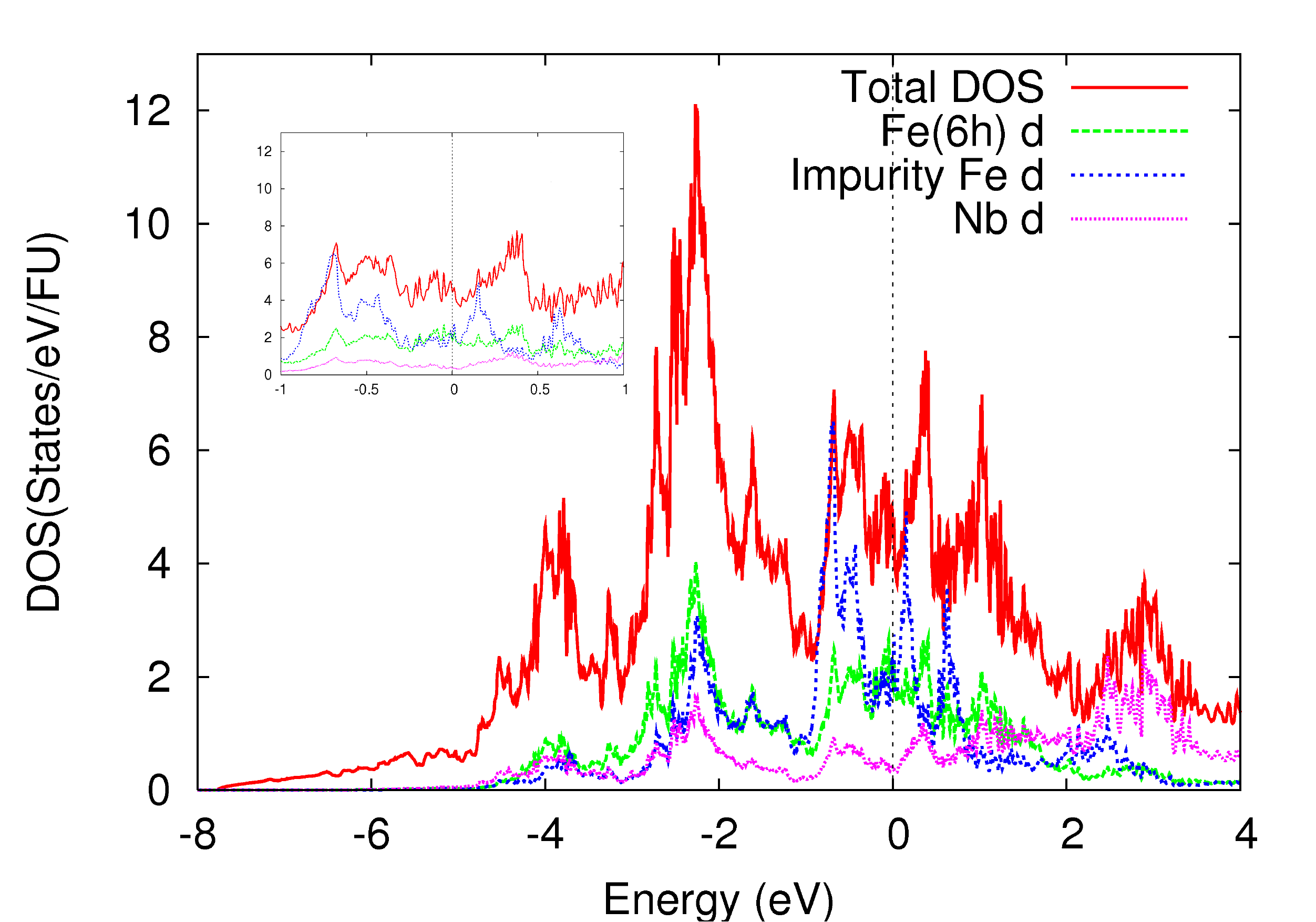}
\end{tabular}
\caption{\label{fig:DOSNbFe2DopedNM} (Color online) Non-magnetic calculated DOS for (a) Nb-rich Nb$_{1.065}$Fe$_{1.935}$ and for (b) Fe-rich Nb$_{0.935}$Fe$_{2.065}$. The insets show the same data on an expanded scale around the Fermi level. The Fermi level is set to 0 eV.}
\end{figure}

In Fig.~\ref{fig:DOSNbFe2DopedNM}(a) we show the calculated density of states for Nb-rich Nb$_{1.065}$Fe$_{1.935}$, with the Nb substituted on the Fe(6\textit{h}) site. Here we find that $N(E_F)=3.8$ eV$^{-1}$FU$^{-1}$. This is a smaller rise than that predicted by the rigid band approximation, but proabably still large enough to satisfy the criterion for the formation of ferromagnetism under Stoner theory. If we inspect the projected density of states for the Nb impurity and an Fe(6\textit{h}) site shown in the same figure, then we can see a mechanism for this result. The Fe(6\textit{h}) site contributes significant structure to the density of states near the Fermi level. As a result when we dope a Nb onto a Fe(6\textit{h}) site, not only are we removing valence electrons from the system, but we are also decreasing the contribution from Fe(6\textit{h}) sites to the density of states at the Fermi level. The importance of the structure of Fe-\textit{d} states has been noted by Inoue \textit{et al.} in tight-binding calculations incorporating impurity atoms\cite{InoueNb1}. Consequently, the density of states at the Fermi level from our doped supercell calculation is lower than that expected in the rigid band approximation.


Next we consider electron doping on the Fe-rich side of the phase diagram of \nbfy. We consider doping corresponding to $y=0.065$ Fe-rich Nb$_{0.935}$Fe$_{2.065}$. This state is formed by substituting one Fe atom for a Nb atom in  a $2 \times2$ supercell and is well within the ferromagnetic region of the experimental phase diagram. However, according to the rigid-band approximation, this doping would actually produce a small fall in the density of states at the Fermi level to $N(E_F)=3.5$ eV$^{-1}$FU$^{-1}$. This is due to the fact that the Fermi level effectively moves higher in energy in Fig.~\ref{fig:DOSNbFe2NM} and as a result the density of states falls deeper into the trough near the Fermi level. Therefore, in the rigid band approximation the Stoner criterion would not be satisfied and ferromagnetism would not result. In Fig.~\ref{fig:DOSNbFe2DopedNM}(b) we show the density of states from our supercell calculation of Nb$_{0.935}$Fe$_{2.065}$. Critically, the density of states at the Fermi level actually rises to $N(E_F)=4.5$ eV$^{-1}$FU$^{-1}$. This is due to the fact, that despite the addition of valence electrons, the Fe dopant contributes increased structure to the density of states at $E_F$. To illustrate this we show in Fig.~\ref{fig:DOSNbFe2DopedNM}(b) the projected density of states for the Fe dopant at the Nb site. We see that, unlike Nb, this Fe dopant contributes a form to the density of states that is very similar to that of an Fe(6\textit{h}) site.

Also, we have made a simple approximation to the disorder by applying a gaussian broadening of width 41meV to our density of states to reflect the decreased lifetime of electrons in the states near the Fermi level due to disorder. This is calculated from the measured low temperature resistivity in a Drude approximation\cite{SoukoulisNb1, TestardiNb1, Moroni1}. In Nb-rich Nb$_{1.065}$Fe$_{1.935}$ we find that $N(E_F)=4.6$ eV$^{-1}$FU$^{-1}$. The reason for the large increase compared to $N(E_F)=3.8$ eV$^{-1}$FU$^{-1}$ found without this broadening may be ascertained from the inset to Fig.~\ref{fig:DOSNbFe2DopedNM}(a). The Fermi level lies in a sharp dip in the density of states and therefore disorder induced broadening is likey to smear out this structure and increase $N(E_F)$. This suggests that disorder may be important to the development of ferromagnetism upon Nb doping. In contrast $N(E_F)$ is found to be insensitive to the broadening for Fe-rich Nb$_{0.935}$Fe$_{2.065}$. The influence of disorder may also explain discrepancies between experimentally determined phase diagrams where some work\cite{Crook1} shows the development of ferromagnetism for lower Nb dopings than in more recent studies\cite{Brando1, Moroni1}. This indicative result invites further work on the description of disorder with approaches such as the Coherent Potential Approximation\cite{SovenNb1, GonisNb1, TurekNb1, PindorNb1}. A further consideration in such work may be how the interaction, $I$, changes with disorder.

These findings illustrate several important points about the evolution of the temperature-doping phase diagram. 1) A rigid-band approximation is not sufficient to understand the evolution to ferromagnetism, particularly for electron doping. 2) The type of dopant will be critical to the character of this phase diagram. This is essentially because the fine structure of this phase diagram is strongly affected by the contributions to the density of states at the Fermi level from Fe-\textit{d} states. 3) The presence of disorder may be important to the development of ferromagnetism upon Nb doping.

We have also undertaken calculations for dopings of $y=\pm0.0325$ using $2 \times2 \times 2$ supercells. These calculations deliver results with the same qualitative conclusions. For example we considered doping corresponding to $y=0.0325$ Fe-rich Nb$_{0.9675}$Fe$_{2.0325}$. This state is formed by substituting one Fe atom for a Nb atom in  a $2 \times2 \times2$ supercell. Applying the rigid-band approximation would result in a fall from $N(E_F)=3.6$ eV$^{-1}$FU$^{-1}$ in stoichiometric \nbf~to $N(E_F)=3.0$ eV$^{-1}$FU$^{-1}$. Our supercell calculation for Fe-rich Nb$_{0.9675}$Fe$_{2.0325}$ indicates $N(E_F)=4.1$ eV$^{-1}$FU$^{-1}$ and supports the formation of ferromagnetism as shown in the experimental phase diagram of Fig.~\ref{fig:DNbFe2}.

\section{\label{sec:Doping}Fermi Surface and Magnetic Instabilities\protect}
To further our understanding of the complex evolution of the magnetic phase diagram with doping we here consider the magnetic interactions active in stoichiometric \nbf, which has experimentally been interpreted as having SDW character. The previous work of Subedi \textit{et al.} has considered the interaction of different magnetic orders within the unit cell\cite{SubediNb1} under the local spin density approximation (LSDA).

\begin{figure*}
\centering
\begin{tabular}{ccc}
(a) Band 81 & (b) Band 82 & (c) Band 83 \\
\includegraphics[width=0.3\textwidth,angle=0]{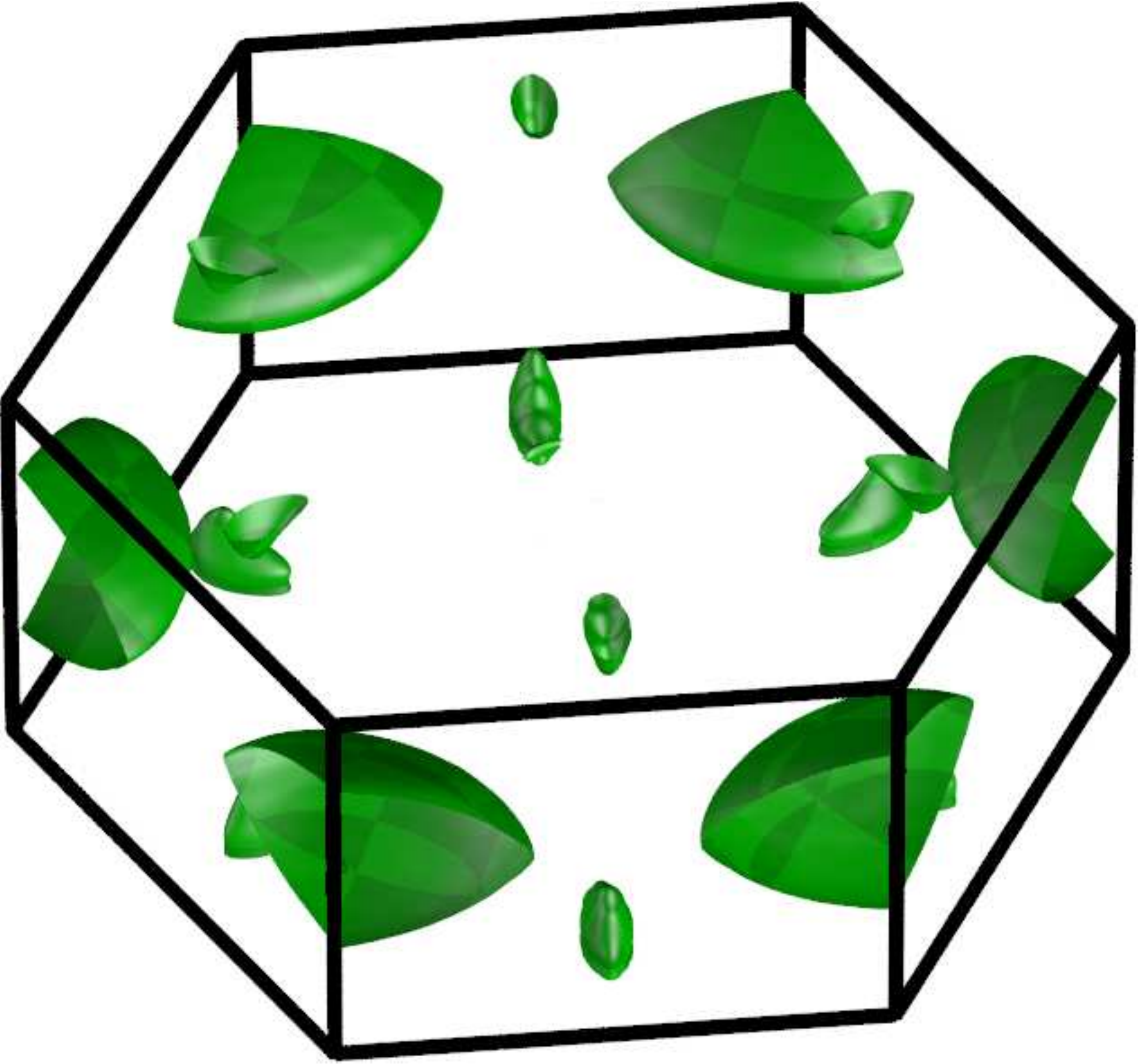} & 
\includegraphics[width=0.3\textwidth,angle=0]{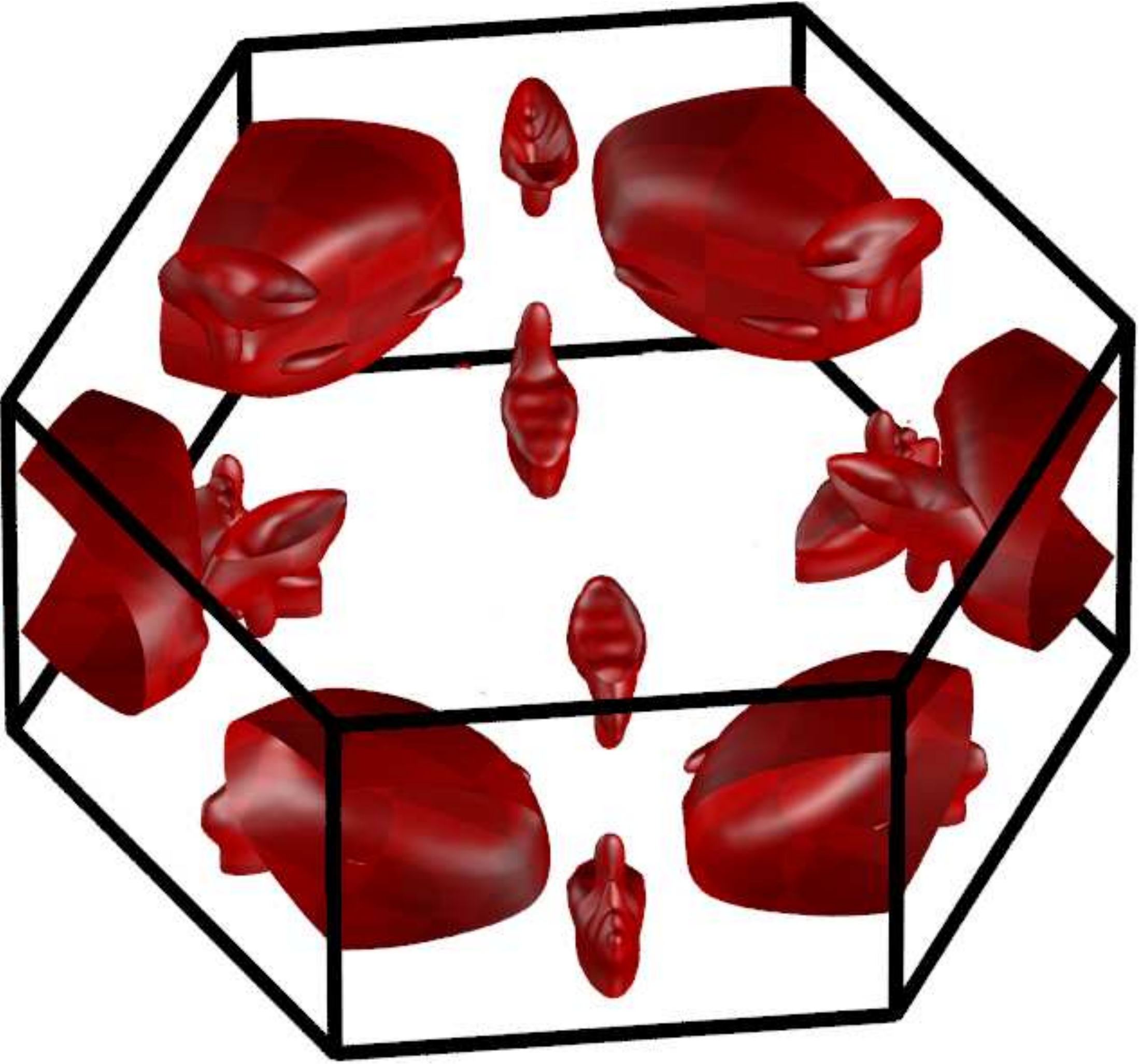} &
 \includegraphics[width=0.3\textwidth,angle=0]{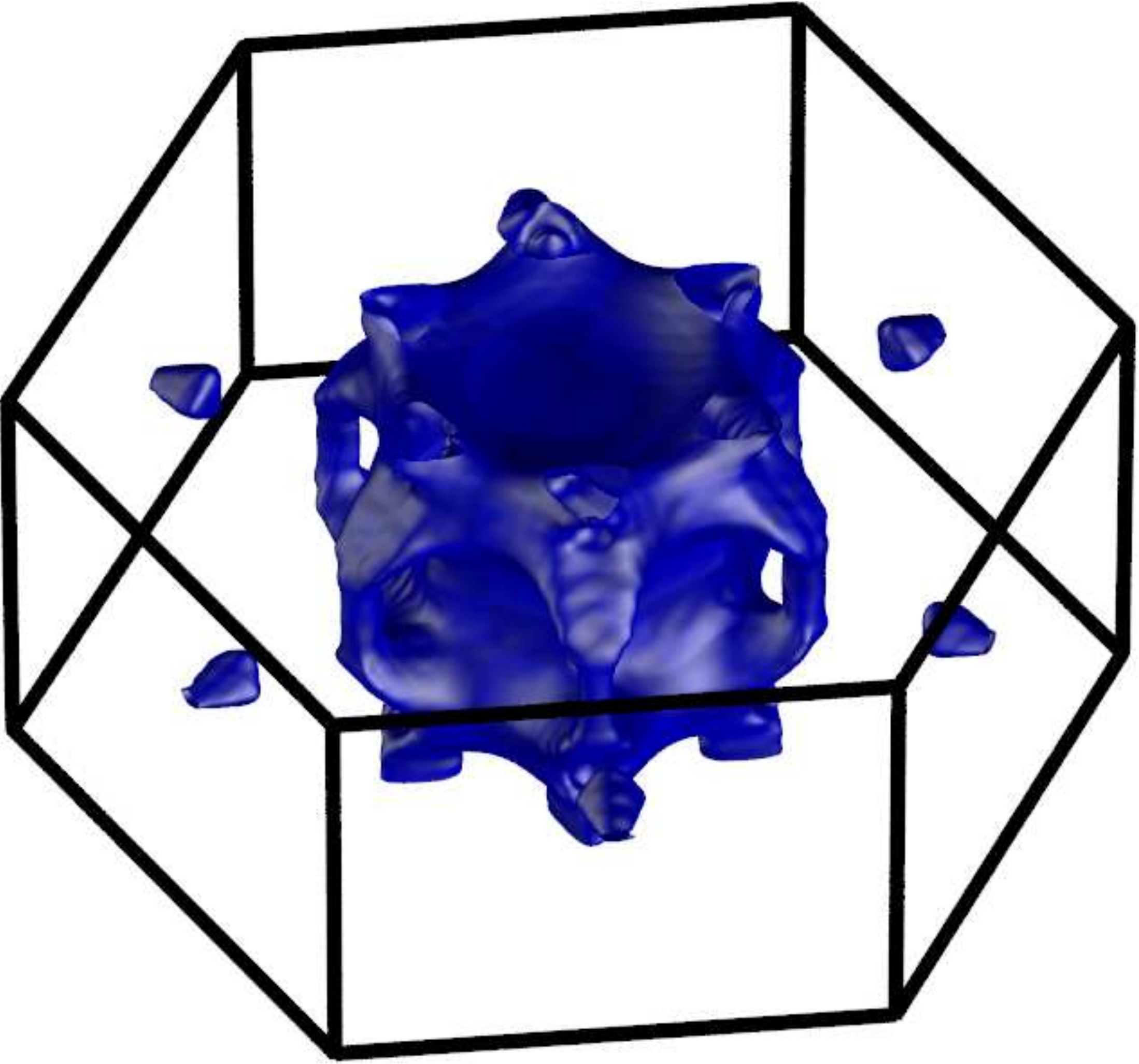} \\
\\
(d) Band 84 & (e) Band 85 & (f) Susceptibility $\chi_{0}(\mathbf{q})$ \\
\includegraphics[width=0.3\textwidth,angle=0]{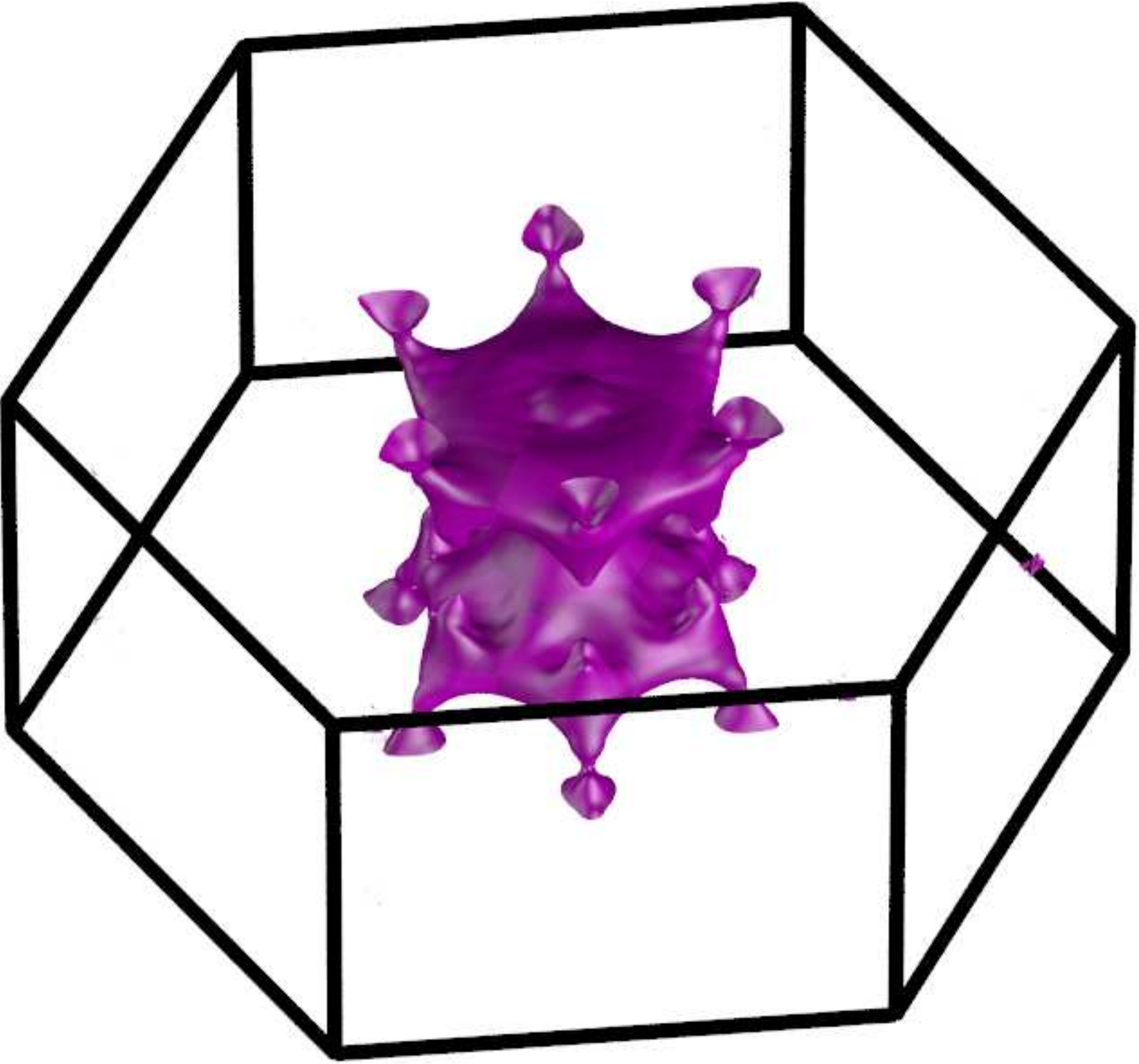} &
\includegraphics[width=0.3\textwidth,angle=0]{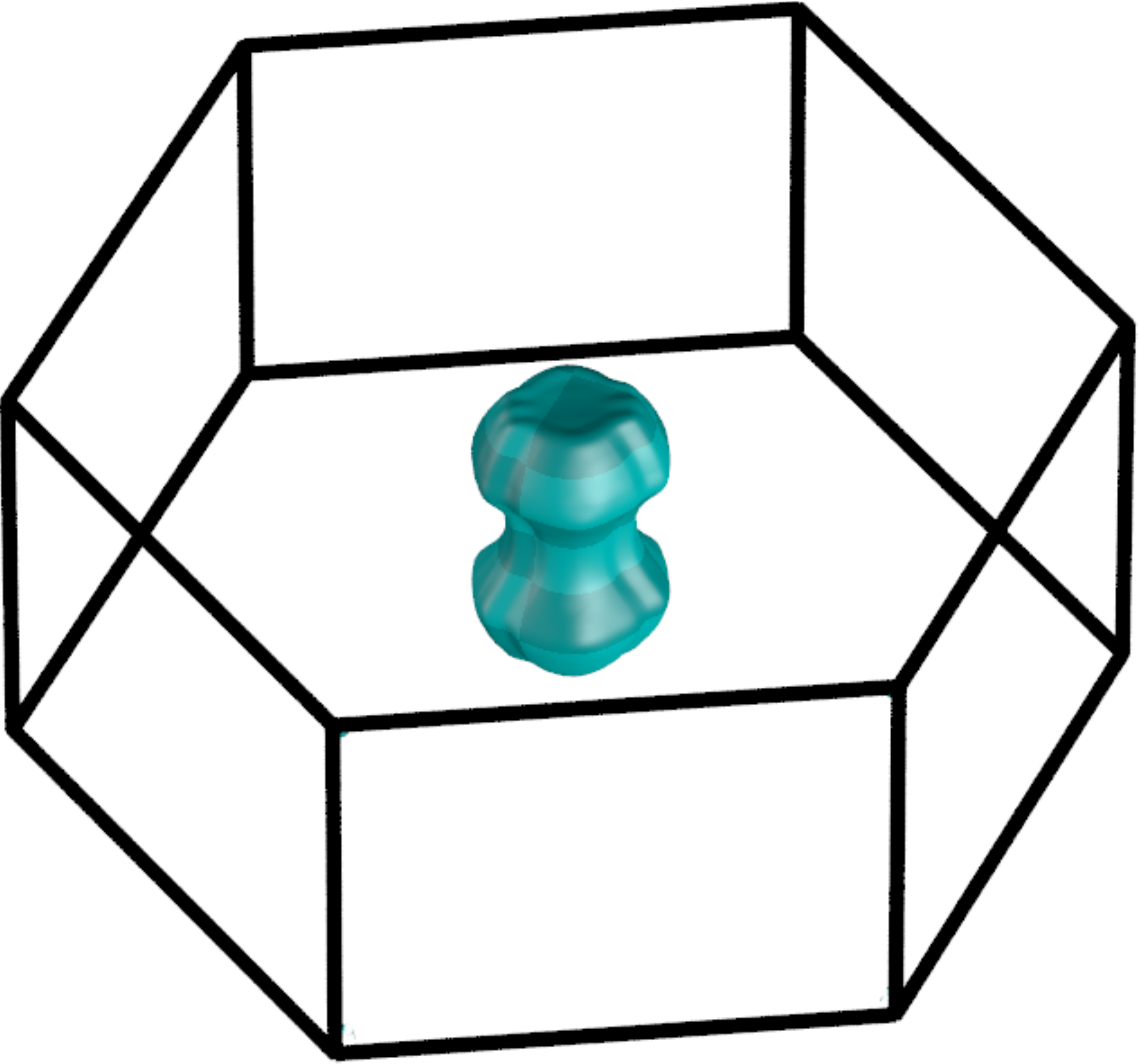} &
\includegraphics[width=0.3\textwidth,angle=0]{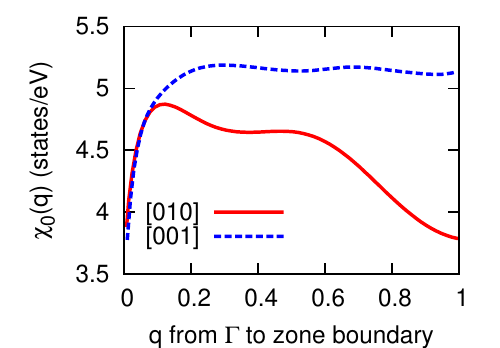}
\\
\end{tabular}
\caption{\label{fig:FSNbFe2} (Color online) In (a)-(e) we show the non-magnetic Fermi surface of \nbf. The flat sections of Fermi surface in bands 81 and 82 show potential for nesting. In (f) we show the non-interacting susceptibility, $\chi_{0}(\mathbf{q})$, for NbFe$_{2}$ along both the [010] and [001] directions. The large flat enhancement in $\chi_{0}(\mathbf{q})$ along the [001] direction indicates the potential importance of large $\mathbf{q}$ magnetic instabilities.
The flat sections of the Fermi surface of the bands 82 and 83 are the dominant contribution to the susceptibility at large $\mathbf{q}$. A similar phenomenon is observed due to the cubic shaped Fermi surface in TiBe$_{2}$\cite{Jeong1}.}
\end{figure*}

The Fermi surface of metallic materials is crucial in the understanding of their properties. In Fig.~\ref{fig:FSNbFe2} we show the calculated non-magnetic Fermi surface for stoichiometric~\nbf~which agrees well with previous work\cite{SubediNb1}. We reproduce this Fermi surface to emphasise the importance of features that may be important to the magnetic instabilities of the system. Five sheets make up this Fermi surface which is strongly 3D. We consider the non-magnetic Fermi surface because we are interested in the magnetic instabilities as we approach the QCP where the static moment and spin splitting of the Fermi surface approach zero.

We proceed by calculating $\chi_{0}(\mathbf{q})$ using the technique of Romberg integration.
\begin{equation}
\mathbf{\chi_{0}(\mathbf{q})} = 
\sum_{n,\mathbf{k}}M^2_{\mathbf{k},\mathbf{k+q}}\left[ \frac{f_{n,\mathbf{k}} - f_{n,\mathbf{k} + \mathbf{q}}}{\varepsilon_{n,\mathbf{k} + \mathbf{q}} - \varepsilon_{n,\mathbf{k}}}
 \right]
\end{equation} 
Here, the sum is over all bands up to 1eV above the Fermi level, $n$, and wave vectors $\mathbf{k}$. $M$ are the matrix elements and $f$ is the Fermi occupation function. No increased lifetime broadening was applied.
In Fig.~\ref{fig:FSNbFe2} (f) we show the susceptibility for $\mathbf{q}$ parallel to $\mathbf{[010]}$ and for $\mathbf{q}$ parallel to $\mathbf{[001]}$. We have included only \textit{intraband} contributions for which we may to a good approximation set the matrix element $M=1$. Both susceptibilities rise steeply away from $\mathbf{q}=0$. Along $\mathbf{[010]}$, the susceptibility peaks at a small wavevector, $\mathbf{q}\approx\mathbf{[0~0.075~0]\pi/b}$, and then falls monotonically until the zone boundary is reached. Along $\mathbf{[001]}$ the susceptibility rises to a higher level than that in-plane $\approx40\%$ above that at $\mathbf{q}=0$. It then levels off to an essentially constant value until the zone boundary is reached. Enhancements at finite $\mathbf{q}$ are consistent with previous work\cite{SubediNb1}. From these calculations it appears likely that finite $\mathbf{q}$ instabilities, particularly along $\mathbf{[001]}$, may be important in this system. 

We note also, that in the absence of spin orbit effects the matrix element for \textit{interband} contributions is $M=0$ at $\mathbf{q}=0$ due to orthogonality, then the inclusion of \textit{interband} contributions will enhance the finite $\mathbf{q}$ instabilities shown in Fig.~\ref{fig:FSNbFe2}(f). 

Inspection of the contributions of each Fermi surface sheet shows that the rising value of $\mathbf{\chi_{0}(\mathbf{q})}$ along $\mathbf{[001]}$ is due largely to bands 82 and 83. This results from the presence of the very flat sections of surface that they possess. These flat sections nest strongly as they are translated by the wavevector $\mathbf{q}$. This behavior is very similar to the large finite $\mathbf{q}$ instability found due to the flat sides of the cube-like Fermi surfaces in metamagnetic TiBe$_2$\cite{Jeong1}.

\begin{figure}
\centering
\begin{tabular}{ccc}
(a) & ~~~~~~~~ & (b) \\
\includegraphics[width=0.4\linewidth,angle=0]{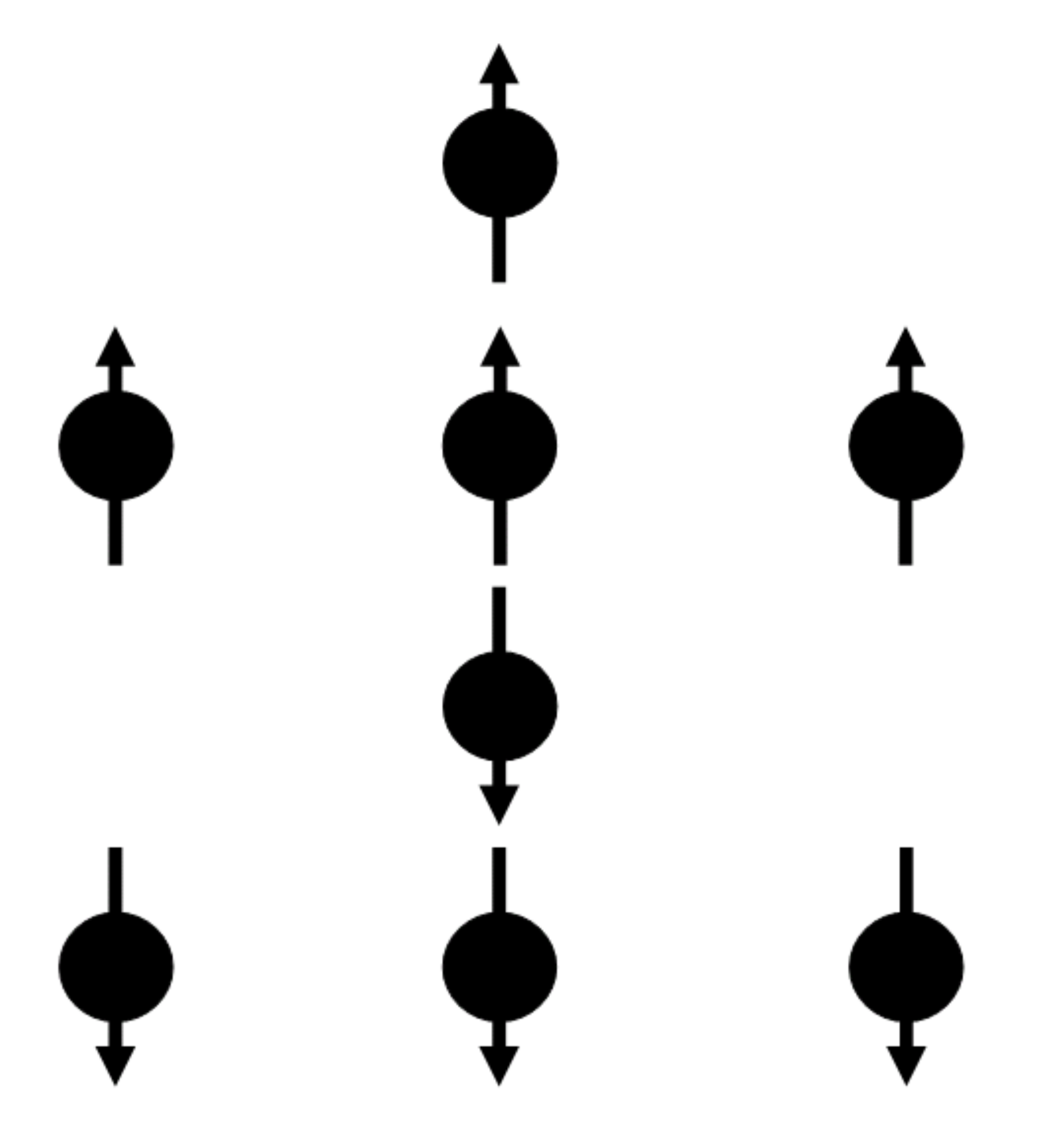}
& ~~~~~~~~ & \includegraphics[width=0.4\linewidth,angle=0]{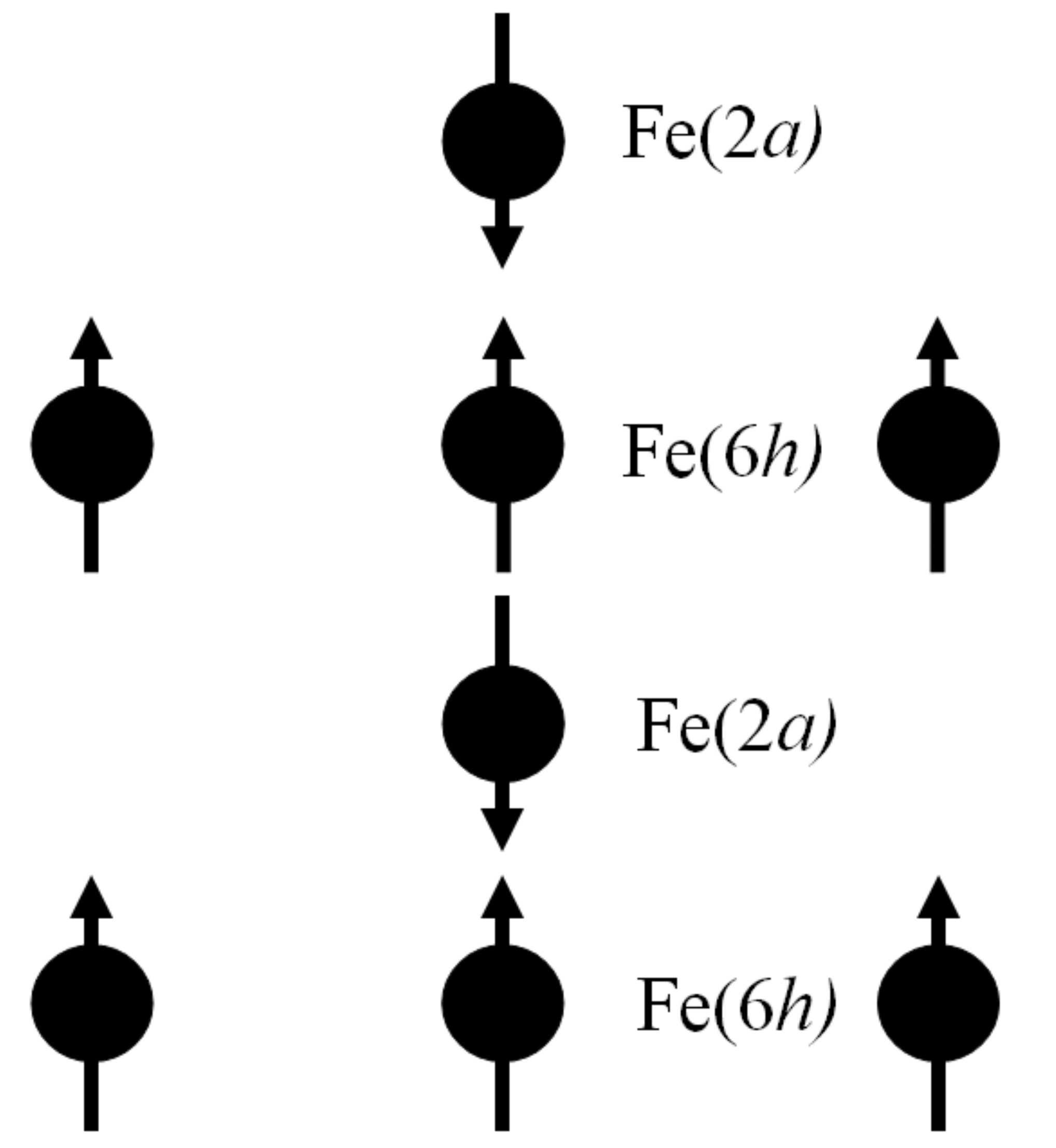}
 \\
\end{tabular}
\caption{\label{fig:spinConfigs} (Color online) Schematic diagram of the spin configuration for (a) AFM2 $\mathbf{q}=0$ antiferromagnetism and (b) AFM3 the lowest energy state found by Subedi \textit{et al.}\cite{SubediNb1}. Only Fe atoms are shown and the 2$a$ and 6$h$ rows are as indicated.}
\end{figure}

We also compare competing magnetic ground states by performing total energy calculations.
Our total energy calculations consider several magnetic states including 1) Non-magnetic, 2) Ferromagnetic,  3) AFM1 which is commensurate along the c-axis as suggested by the calculation of $\chi_{0}(\mathbf{q})$, 4) AFM2 which is $\mathbf{q}=0$ antiferromagnetism of the type found in the C14 Laves material TiFe$_{2}$ and 5) AFM3 which is the lowest energy state found by Subedi \textit{et al.}\cite{SubediNb1}. We show the spin configurations associated with some of these orders in Fig.~\ref{fig:spinConfigs}.

For the energy evaluation we have used 10,000 k-points in the Brillouin zone and the GGA-PBE exchange correlation functional.
As shown in Table~\ref{tab:NbFe2Energetics} we find that the lowest energy state corresponds to $\mathbf{q}=0$ antiferromagnetism of the type found experimentally in the C14 Laves material TiFe$_{2}$\cite{BrownTh1}. However, our results indicate that the Fe(2$a$) possesses a moment, while the experiments on TiFe$_{2}$ indicate no ordered moment. In this calculation the second lowest energy state is then the ferrimagnetic state found to have the lowest energy by Subedi \textit{et al.}\cite{SubediNb1}. The use of the LSDA correlation functional in that study results in a different configuration of spins for the lowest energy state. 

For the SDW order AFM1 corresponding to the peak in $\chi_{0}(\mathbf{q})$ along the c-axis we find a lower energy than that obtained for ferromagnetism, but not lower than the configurations within the unit cell. The ground state calculation for these SDW configurations also produces large moments and suggests that the energy gain is again dominated not by the Fermi surface nesting, but by the Hund's coupling.

\begin{table}
\caption{\label{tab:NbFe2Energetics} Comparative energetics of magnetic states. For the ferromagnetic
state the total magnetic moment is 2.34$\mu_{B}$ per formula unit. In the ferromagnetic state the Nb is
found to also carry a small moment $\approx0.2\mu_{B}$ which is antiferromagnetically coupled to the Fe
moments.}
\begin{ruledtabular}
\begin{tabular}{cc|cc}
  & Energy (meV/FU) & Fe(2\textit{a}) $\mu$($\mu_{B}$) & Fe(6\textit{h}) $\mu$($\mu_{B}$) \\
\hline
$NM$ & 0 & 0 & 0\\
$FM$ & -84 & 1.1 & 1.5\\
$AFM1$ & -96 & 1.6 & 1.4\\
$AFM2$ & -104 & 1.6 & 1.5\\
$AFM3$ & -97 & 1.8 & 0.98\\
\end{tabular}
\end{ruledtabular}
\end{table}


Importantly, the ordered moment below the SDW transition
($ < 0.02\mu_{B}$) represents only a small fraction of the fluctuating
moment\cite{Brando1}. Magnetic order is then expected to have a
minor effect on the magnetic fluctuation spectrum and
hence on the low temperature properties, in particular the non-Fermi liquid low temperature exponents. This is evidenced by the small size of the heat capacity anomaly at
$T_{N}\approx 10 K$ even in stoichiometric \nbf. Therefore, the presence of finite $\mathbf{q}$ magnetic fluctuations may have an important role in producing the non-Fermi liquid properties. \\

From spin fluctuation theory, it is known that at an antiferromagnetic ($\mathbf{q}\neq 0$) quantum critical point the expected temperature dependence of the resistivity is $\rho \propto T^{3/2}$ in the presence of sufficiently strong quenched disorder\cite{HlubinaTh1}, while the leading order term in the specific heat is linear in $T$. In contrast at a ferromagnetic critical point ($\mathbf{q}= 0$), $\rho \propto T^{5/3}$ and the specific heat diverges logarithmically. As shown in Table~\ref{tab:SpinTheory}, experiments on \nbf~ at its quantum critical point might suggest the presence of both ferromagnetic and antiferromagnetic spin fluctuations if we were able to treat these modes separately. Alternatively, \nbf~may be entering into a new regime such as that found in the high pressure region of the helimagnet MnSi\cite{PfleidererTh1}. It is possible that the presence of $\mathbf{q}=0$ and finite $\mathbf{q}$ instabilities in this material may combine to produce the unusual non-Fermi liquid behavior. The prominence of such instabilities in \nbf~is confirmed by both our calculation of the susceptibility in Fig.~\ref{fig:FSNbFe2}(f) and the total energy calculations in Table~\ref{tab:NbFe2Energetics}.

\begin{table}
\caption{\label{tab:SpinTheory} Results from spin fluctuation theory\cite{Moriya1} in 3D shown alongside the experimental results for \nbf~at its quantum critical point\cite{Brando1}. Results are shown for the standard Fermi liquid (FL), a system  near a finite $\mathbf{q}$ antiferromagnetic instability (AFM) and 
a system  near a ferromagnetic instability (FM).}
\begin{ruledtabular}
\begin{tabular}{ccccc}
Property & FL & AFM & FM & NbFe2 \\
\hline
Specific Heat (C) & $T$ & $T$ & $T\ln(T^{*}/T)$ & $T\ln(T^{*}/T)$\\
Resistivity ($\rho$) & $T^{2}$ & $T^{3/2}$ & $T^{5/3}$ & $T^{3/2}$
\end{tabular}
\end{ruledtabular}
\end{table}

The interplay of Fermi surface nesting along with other magnetic couplings in this system is shared with the Fe-based superconductors where nesting is
important along with other band structure effects. The competition of these effects has been shown by Yildirim\cite{YildirimNb1} to produce the structural phase transition in the Fe-based superconductors. In the \nbf~system they may produce intriguing non-Fermi liquid properties.

\section{\label{sec:Conclusions}Summary and Discussion\protect}
The bonding structure of the Fe-cages in \nbf~is likely to be instrumental in producing the magnetic interations that dominate the formation of its magnetic ground state. This bonding produces inequivalent character in the kagome triangles in the Fe(6\textit{h}) layers.

We demonstrate that non-rigid band effects must be considered in order to understand the evolution of the ground state as a function of doping. The contribution of Fe-\textit{d} states to the density of states at the Fermi level is shown to be critical to the production of a Stoner instability upon electron doping with Fe impurities. Therefore, it is expected that the nature of the electron dopant will be crucial to the evolution of the phase diagram. Also, the presence of disorder is likely to be important to the formation of ferromagnetism upon Nb doping.

We find that \nbf~is close to both small and large $\mathbf{q}$ magnetic instabilities. The non-interacting susceptibility indicates the potential importance of $\mathbf{q}\neq 0$ instabilities. The electronic structure is highly three dimensional, but the possibilities for nesting of quasi-1D sections of the Fermi surface may provide crucial magnetic instabilities.  Total energy calculations indicate that spin configurations within the unit cell produce the lowest energy state.

The unusual non-Fermi liquid critical exponents of \nbf~may arise from the combination of $\mathbf{q}\neq 0$ and $\mathbf{q}=  0$ instabilities. Further investigation as to how these instabilities may interplay will be necessary both theoretically and experimentally.
Furthermore, the presence of such a large phase space for fluctuations is likely to complicate the formation of superconducting Cooper pairs in either the spin singlet or triplet channels.\\

The authors would like to acknowledge useful discussions with J.B. Staunton.

\bibliography{Srv3}

\end{document}